\documentclass[11pt]{article}
\usepackage{epsfig,psfig,times,alltt,xspace,subfigure,graphicx,latexsym}

\newcommand{\topseplen}{0ex} %
\newcommand{\crunchbegin}{\addtolength{\topsep}{\topseplen}}
\newcommand{\crunchend}{}   %

\newcommand{\condBox}{{\raggedleft \mbox{$ \Box $}}}
\newenvironment{example}{\begin{ex} \nopagebreak \crunchbegin \rm \begin{rm}}{\condBox \end{rm} \crunchend \end{ex}}
\newtheorem{ex}{Example}[section]

\newcommand{\eg}{e.g.,\xspace}
\newcommand{\ie}{i.e.,\xspace}
\newcommand{\pqueue}{\emph{pqueue}\xspace}
\newcommand{\MinDiv}{\emph{MinDiv}\xspace}

\begin{document}

\pagestyle{plain}
\setcounter{page}{1}

\title{Providing Diversity in K-Nearest Neighbor Query Results}

\author{Anoop Jain \hspace*{0.2in} Parag Sarda \hspace*{0.2in} Jayant R. Haritsa\\
Department of Computer Science \& Automation \\
Indian Institute of Science, Bangalore 560012, INDIA\\
\{anoop, parag, haritsa\} @dsl.serc.iisc.ernet.in}

\date{}
\maketitle

\begin{abstract}
Given a point query Q in multi-dimensional space, K-Nearest Neighbor (KNN)
queries return the K closest answers according to given distance metric in the
database with respect to Q. In this scenario, it is possible that a majority of
the answers may be very similar to some other, especially when the data has
clusters. For a variety of applications, such homogeneous result sets may not
add value to the user.  In this paper, we consider the problem of providing
diversity in the results of KNN queries, that is, to produce the closest result
set such that each answer is sufficiently different from the rest. We first
propose a user-tunable definition of diversity, and then present an algorithm,
called MOTLEY, for producing a diverse result set as per this definition.
Through a detailed experimental evaluation on real and synthetic data, we show
that MOTLEY can produce diverse result sets by reading only a small fraction of
the tuples in the database.  Further, it imposes no additional overhead on the
evaluation of traditional KNN queries, thereby providing a seamless interface
between diversity and distance.
\end{abstract}

\section{Introduction}

Over the last few years, there has been considerable interest in
the database community with regard to supporting K-Nearest
Neighbor (KNN) queries.  The general model of a KNN query is that
the user gives a point query in multidimensional space and a
distance metric for measuring distances between points in this
space. The system is then expected to find, with regard to this
metric, the K closest answers in the database from the query
point. Typical distance metrics include Euclidean distance,
Manhattan distance, etc. 

\begin{example}
\label{example}
Consider the situation where the Athens tourist office maintains
the relation {\sf RESTAURANT (Name,Speciality,Rating,Expense)},
where Name is the name of the restaurant; Speciality indicates
the food type (Greek, Chinese, Indian, etc.);  Rating is an
integer between 1 to 5 indicating restaurant quality; and,
Expense is the typical expected expense per person.  In this
scenario, a visitor to Athens may wish to submit the following
KNN query (using the SQL-like notation of~\cite{TOPK}) to have a
choice of three mid-range restaurants where she can have dinner
for a expense of around 50 Euros.

{\small
\begin{verbatim}
  SELECT * FROM RESTAURANT
  WHERE Rating=3 and Expense=50 
  ORDER 3 BY Euclidean 
\end{verbatim}
}
\end{example}

In practice, databases often have their data clumped together in clusters
-- for example, there could be several Greek restaurants which come close
to the specified values -- in fact, it is even possible that there are
exact \emph{duplicates} (w.r.t. the Rating and Expense attributes).  In
such a situation, returning three very similar answers (e.g.
\emph{\{Parthenon-1,3,54\}}, \emph{\{Parthenon-2,3,45\}}, and
\emph{\{Parthenon-3, 3, 60\}})  may not add much value to the user.
Instead, she might be better served by being told, in addition to
\emph{\{Parthenon-1,3,54\}} about a Chinese restaurant \emph{\{Hunan,2,40\}}
and an Indian restaurant \emph{\{Taj,4,60\}}, which would provide a viable
set of choices to plan her dinner.

That is, the user would like to have not just the closest set of answers,
but the closest \emph{diverse} set of answers (an oft-repeated quote from
Montaigne, the sixteenth century French writer, is ``The most universal
quality is diversity''~\cite{mont})

\begin{figure}[ht]
   \centerline {\includegraphics[height=5cm,width=12cm]{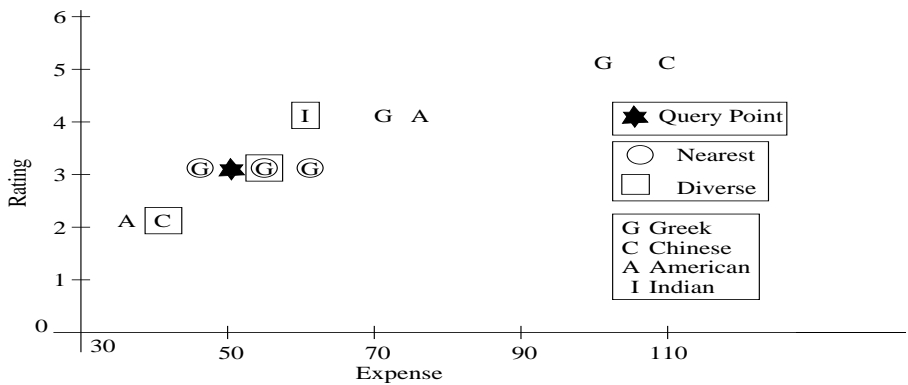}}
   \caption{Diversity Example}
   \label{figure:toy}
\end{figure}

To clarify the above, consider Figure~\ref{figure:toy}, which shows a
sample distribution of data points in the RESTAURANT database, and a
query point supplied by the user.  In this scenario, a traditional KNN
approach would return the answers shown by the circles, all of which are
very similar (returning all Greek restaurants).  What we need however
is to produce the answers shown by the rectangles, representing a close
but more heterogeneous result set (Greek, Indian and Chinese restaurants).

\subsection{The KNDN Problem}
Based on the above motivation, we consider in this paper the problem of
providing diversity in the results of KNN queries, that is, to produce
the closest result set such that each answer is sufficiently diverse
from the rest. We hereafter refer to this as the \emph{K-Nearest
Diverse Neighbor (KNDN)} problem, which to the best of our knowledge
has not been previously investigated in the literature (we explain
in Section~\ref{section:related} as to why the KNDN problem cannot be
handled by traditional clustering techniques).

An immediate question that arises is how to define diversity.  This is
obviously a user-dependent choice. We address the issue by providing a
tunable definition that can be set with a single parameter, \MinDiv,
by the user. \MinDiv values range over [0,1] and specify the minimum
diversity that should exist between \emph{any pair} of answers in the
result set. Note that this is similar to the user specifying $minsup$
(minimum support) and $minconf$ (minimum confidence) to determine what
constitutes an interesting correlation in association rule mining.
Setting \MinDiv to zero results in the traditional KNN query, whereas
higher values give more and more importance to diversity at the expense
of distance. In our framework, a sample query looks like

{\small
\begin{verbatim}
  SELECT * FROM RESTAURANT 
  WHERE Rating=3 and Expense=50
  ORDER 3 BY Euclidean WITH MinDiv=0.1 ON (Speciality)
\end{verbatim}
}

\noindent
where \emph{Speciality} is the attribute on which diversity is calculated, and the
goal is to produce the closest result set that obeys the diversity
constraints specified by the user.  

Unfortunately, as we will explain later in the paper, finding the
optimal result set for KNDN queries is an \emph{NP-complete problem} 
in general, and is computationally extremely expensive even for fixed K,
making it infeasible in practice.  Therefore, we present an alternative
online algorithm, called {\bf MOTLEY}\footnote{Motley: A collection
containing a variety of things}, for producing a sufficiently diverse
and close result set.  Motley adopts a greedy heuristic and assumes the
existence of a multidimensional index with containment property, such as
the R-tree, which is natively available in today's commercial database
systems~\cite{ORACLE}.  The R-tree index supports a ``distance browsing''
mechanism proposed in  \cite{RTR} which allows us to efficiently access
database points in increasing order of distance from the query point.
A pruning technique is incorporated in Motley to minimize the R-tree
processing and the number of database tuples that are examined.

Through a detailed experimental evaluation on real and synthetic data,
we show that Motley can produce a diverse result set by reading only
a small fraction of the tuples in the database.  Further, the quality
of its result set is very close to that provided by an off-line optimal
algorithm. Finally, it can also evaluate traditional KNN queries without
any added cost, thereby providing a \emph{seamless interface between the
orthogonal concepts of diversity and distance.}  While the algorithms
and experiments presented in this paper are for databases where the
diversity attributes are numeric, we also discuss in detail how the
Motley algorithm can be extended to handle \emph{categorical attributes}.

\subsection{Organization}
The remainder of this paper is organized as follows: The basic
concepts underlying our problem formulation are described in
Section~\ref{section:concepts}.  The Motley algorithm is
presented in Section~\ref{section:motley}. The performance model
and the experimental results are highlighted in
Section~\ref{section:experiments}.  Related work on nearest
neighbor queries is overviewed in Section~\ref{section:related}.
Finally, in Section~\ref{section:conclusions}, we summarize the
conclusions of our study and outline future avenues to explore.

\section{Basic Concepts and Problem Formulation
\label{section:concepts}}

We assume that the database is composed of N tuples over a D-dimensional
space $(d_1,d_2,\ldots,d_D)$.  For ease of exposition, we assume for
now that the domains of all attributes are numeric and normalized to
the range [0,1]. Later, in Section~\ref{section:categorical}, we discuss
how to handle categorical attributes.

The user specifies a point query Q over an M-sized subset of these
attributes $(d_{q_1},d_{q_2}, \dots ,d_{q_M}), M \le D$. We refer to
these attributes as ``point attributes'' and the space formed by these
attributes as \emph{spatial-space}. The user also specifies $K$, the
number of desired answers, and a L-sized subset of attributes on which
she would like to have diversity $(d_{v_1},d_{v_2}, \dots ,d_{v_L}), L
\le D$. We refer to these attributes as ``diversity attributes'' and the
associated space as \emph{diversity-space}. Note that the choice of the
diversity attributes is \emph{orthogonal} to the choice of the query's
point attributes.  Referring back to Example~\ref{example}, $D= 4, M=2,
L=1$, $d_{q_1} = Rating, d_{q_2} = Expense$ and $d_{v_1} = Speciality$.

For simplicity, we assume in the following discussion that all dimensions
are equivalent in that the user has no special affinity for one dimension
or the other -- the extension to the biased case is straightforward,
as explained in \cite{tech-report}.

\subsection{Result Diversity} 
For the KNDN problem, diversity is defined with regard to a \emph{pair} of
points and is evaluated with respect to the diversity attributes mentioned
in the query specification. Specifically, given points $P_1$ and $P_2$, and
$V(Q)$, the set of diversity attributes in query $Q$,
the function $DIV(P_1, P_2, V(Q))$ returns true if
points $P_1$ and $P_2$ are diverse with respect to each other on the specified
dimensions. A sample DIV function is given in the following subsection.

Given that there are N points in the database and that we need to select
K points for the result set, there are $^NC_K$ possible choices. An
additional constraint is that we require all points in the result set
to be diverse with respect to each other. That is, given a result set
$\cal R$ with points $R_1, R_2, \ldots, R_K$, we require $DIV(R_i,R_j,V(Q))
= \mbox{ true }\; \forall\; i,j$ such that $i \neq j$ and $1 \leq
i,j \leq K$.  We call such a result set to be \emph{fully diverse}.
As explained later, there may occur situations wherein \emph{no} fully
diverse result set is feasible, in which case we have to settle for a
\emph{partially diverse} result set.

\subsection{Diversity Function}
While what exactly constitutes diversity is obviously a user-specific
perception, we describe here a diversity function that, in our opinion,
reflects what would be typically expected in practice. We hasten to
add here that the specific choice of diversity function does not affect the
algorithms presented subsequently in the paper.

Our computation of the diversity between two points $P_1$ and $P_2$,
is based on the classical \emph{Gower coefficient}~\cite{GOWER},
wherein the difference between two points is defined as a weighted
average of the respective attribute differences.  Specically, we first
compute the differences between the attribute values of these two
points in \emph{diversity-space}, and then sequence these differences
in \emph{decreasing} order of their values (recall that the values on
all dimensions are normalized to [0,1]) and label them as $({\delta}_1,
{\delta}_2, \ldots, {\delta}_L)$.
\begin{example}
Consider a pair of points $P_1, P_2$, and a query $Q$ with three
diversity attributes, let the associated differences be 0.4, 0.3 and 0.5. In
this case, we produce ${\delta}_1=0.5, {\delta}_2=0.4, {\delta}_3=0.3$.
\end{example}
Now, we calculate $divdist$, the diversity distance of points $P_1$
and $P_2$ with respect to query $Q$ as
\begin{equation}
\label{eq:basicdiv}
  divdist(P_1,P_2,V(Q)) = \sum _{j=1}^{L} (W_{j} \times{\delta}_j )
\end{equation}
\noindent
where the $W_j$'s are weighting factors for the differences.
Since all ${\delta}_j$'s are in the range [0,1], and by virtue of
the $W_j$ assignment policy discussed below, diversity values are
also bounded in the range [0,1].

The assignment of the weights is based on the heuristic that \emph{larger}
weights should be assigned to the larger differences. That is, in
Equation~\ref{eq:basicdiv}, we need to ensure that $W_i \ge W_j$ if $i
< j$. The rationale for this assignment is as follows: Consider the
case where point $P_1$ has values (0.2, 0.2, 0.3), point $P_2$ has
values (0.19, 0.19, 0.29) and point $P_3$ has values (0.2, 0.2, 0.27).
Consider the diversity of $P_1$ with respect to $P_2$ and $P_3$. While the
aggregate difference is the same in both cases, yet intuitively we can see
that the pair ($P_1$, $P_2$) is more homogeneous as compared to the pair
($P_1$, $P_3$). This is because $P_1$ and $P_3$ differ considerably on the
third attribute as compared to the corresponding differences between $P_1$
and $P_2$.  That is, a pair of points that have higher variance in their
attribute differences appear more diverse than those with lower variance.

Now consider the case where $P_3$ has value (0.2, 0.2, 0.28).  Here,
although the aggregate ${\delta}_j$ is higher for the pair ($P_1$,
$P_2$), yet again it is pair ($P_1$,$P_3$) that appears more diverse
since its difference on the third attribute is larger than any of the
individual differences in pair ($P_1$,$P_2$).

Based on the above discussion, the weighting function should have
the following properties: Firstly, all weights should be positive,
since differences in any dimension should never decrease the
diversity. Second, the sum of the weights should add up to 1
(\ie $\sum_{j=1}^{M} W_{j} = 1$) to ensure that $divdist$ values
are normalized to the [0,1] range. Finally, the weights should be
\emph{monotonically decaying} ($W_i \ge W_j$ if $i < j$ ) to reflect the
preference given to larger differences.

\begin{example}
A candidate weighting function that obeys the above requirements is
the following:
\begin{equation}
\label{eq:weights}
W_{j}  = \frac{a^{j-1}\times(1-a)} {1-a^{M}} \; \; (1 \leq j \leq M)  
\end{equation}
\noindent where $a$ is a tunable parameter over the range (0,1).  Note that
this function implements a \emph{geometric} decay, with the parameter `$a$'
determining the rate of decay. Values of $a$ that
are close to 0 result in faster decay, whereas values close to 1 result in slow
decay. When the value of $a$ is nearly 0, almost all weight is given to maximum
difference \ie $W_{1} \simeq 1$, modeling the \emph{MAX} metric.  Conversely,
when $a$ is nearly 1, all attributes are given similar weights, modeling a
\emph{scaled Manhattan} metric. Figure~\ref{figure:div} shows, for different
values of the parameter `\emph{a}' in Equation~\ref{eq:weights}, the locus of
points which have a diversity of 0.1 with respect to the origin (0,0) in 
two-dimensional diversity space.

\begin{figure}[h]
   \centerline{\includegraphics[height=6cm,width=9cm]{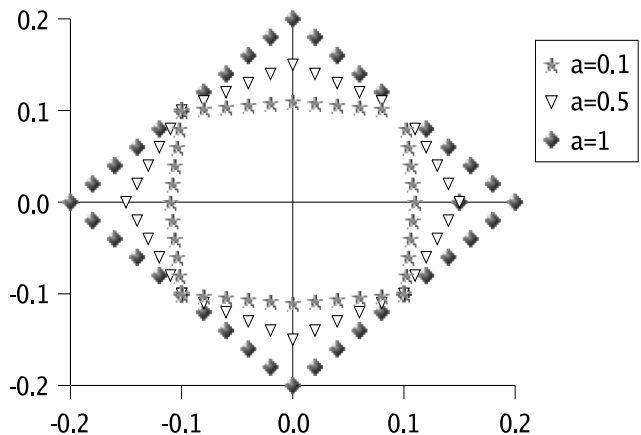}}
   \caption{Points having diversity of 0.1 with respect to (0, 0)}
   \label{figure:div}
\end{figure}

\end{example}

\subsubsection{Directional Diversity}
In the above discussion, differences in either \emph{direction} of a
diversity dimension were considered equivalent -- however,
there may be cases where the user may prefer a given direction.
For example, when purchasing a product, the user may prefer
diversity with respect to lower prices rather than 
higher prices.  Extension of the above formulation to handle
such situations is described in \cite{tech-report}.

\subsubsection{Minimum Diversity Threshold}
We assume that the user provides a quantitative notion of the minimum
diversity that she expects in the result set through a threshold
parameter \MinDiv that ranges between [0,1]. (This is similar
to the setting of \emph{minsup} and \emph{minconf} in association rule
mining to determine what constitutes interesting correlations.) Given
this threshold setting, we say that two points are diverse if the diversity
between them is greater than \MinDiv. That is,
\begin{eqnarray*}
DIV(P_1, P_2, V(Q)) &=& \mbox{ true }\hspace*{10mm} 
	\mbox{ if } divdist(P_1,P_2,V(Q)) \ge MinDiv\\
DIV(P_1, P_2, V(Q)) &=& \mbox{ false }\hspace*{10mm} 
	\mbox{ otherwise }
\end{eqnarray*}

We can provide a \emph{physical} interpretation of the \MinDiv value: If two
points are deemed to be diverse, then these two points have a difference of at
least \MinDiv on one diversity dimension. For example, a
\MinDiv of 0.1 means that any pair of points in the result set differ 
in at least one diversity dimension by at least 10\% of the associated domain size.
This physical interpretation can help guide the user in determining the
appropriate setting of \MinDiv. In practice, we expect that users would choose
\MinDiv values in the range of 0 to 0.2. 
As a final point,  note that with the above formulation, the $DIV$ function
is \emph{symmetric} with respect to the point pair $\{P_1, P_2\}$.  However, it is
\emph{not transitive} in that even if $DIV(P_1,P_2,V(Q))$ and $DIV (P_2,P_3,V(Q))$
are both true, it does not imply that $DIV(P_1,P_3,V(Q))$ is true.

\subsection{Integrating Diversity and Distance}
After applying the diversity constraint, there may be a \emph{variety}
of fully diverse sets that are feasible. We now bring in the notion
of distance from the query point to make a selection between these
sets. That is, we would prefer the fully diverse result set whose
points lie \emph{closest} to the query point Q. Viewed abstractly, we have
a \emph{two-level} scoring function: The first level chooses candidate result sets
based on diversity constraints and the second level selects the result set which 
is spatially closest to the query point. 

Let function $spatialdist(P, Q)$ calculate the spatial distance of point $P$ from
query point $Q$ (recall that this distance is computed with regard to the point attributes
on which $Q$ is specified).  The choice of spatialdist function 
is based on the user
specification and could be any monotonically increasing distance function such
as Euclidean, Manhattan, etc.  We combine distances of all points in a set into
a single value using an aggregate function $Agg$ which captures the overall
distance of the set from $Q$.  While a variety of aggregate functions are
possible, the choice is constrained by the fact that the aggregate function
should ensure that as the points in the set move farther away from the query,
the distance of the set should also increase correspondingly.  Sample aggregate
functions which obey this constraint include the Arithmetic,
Geometric, and Harmonic Means.

We use the reciprocal of the aggregate of the distances of the points
from the query point to determine the score of the set.  That is,
given a query $Q$ and a candidate fully diverse result set $\cal R$,
the score of $R$ w.r.t. $Q$ is computed as
\begin{equation} 
Score({\cal R},Q) = \frac{1}{{Agg}(spatialdist(Q,R_1),\dots,spatialdist(Q,R_K))}
\label{eqn:score}
\end{equation}

\subsection{Problem Formulation
\label{section:problem_formulation}}

\noindent
In summary, our problem formulation is as follows:

\emph{ Given a point query Q on a D-dimensional database, a
desired result cardinality of K, and a
MinDiv threshold, the goal of the K-Nearest Diverse
Neighbor (KNDN) problem is to find the set of K diverse tuples in the
database, whose score, as per Equation~\ref{eqn:score}, is the
maximum, after including the nearest tuple to Q in the result
set.  }

The requirement that the nearest point to the user's query point should
\emph{always} form part of the result set is because this point, in a
sense, \emph{best fits} the user's query.  Therefore, the result sets are
differentiated based on their remaining $K-1$ choices since point $R_1$
is fixed. Further, the nearest point $R_1$ serves to seed the result
set since the diversity function is meaningful only for a \emph{pair}
of points.  

An important point to note here is that the KNDN problem reduces to the
traditional KNN problem when \MinDiv is set to zero.

\subsection{Problem Complexity
\label{section:problem_complexity}}
Finding the optimal result set for the KNDN problem is computationally hard. We
can establish this by mapping KNDN to the well known \emph{independent set
problem}~\cite{MG} which is NP-complete.  The mapping is achieved by forming a
graph corresponding to the dataset in the following manner: Each tuple in the
dataset forms a node in the graph and a edge is added between two nodes if the
diversity between the associated tuples is less than \MinDiv.  Now any
independent set (subgraph in which no two nodes are connected) of size $K$ in
this graph represents a fully diverse set of K tuples.  But finding any
independent set, let alone the optimal independent set, is itself
computationally hard. The straight forward method which checks for all possible
$^NC_K$ sets has $O(N^K)$ running time complexity, which means that even for
fixed $K$, the method is not practically feasible due to high computational
costs.  Tractable solutions to the independent set problem have been recently
proposed~\cite{MG}, but they require the graph to be sparse and all nodes to have a
bounded small degree. In our world, this translates to requiring that all the
clusters in diversity space should be small in size.  But, this may not be
typically true for the datasets that we encounter in practice and
therefore, these solutions may not be applicable in our environment.

\section{The MOTLEY Algorithm} 
\label{section:motley}

We move on, in this section, to present the Motley algorithm, our online
solution technique for the KNDN problem.  Since identifying the optimal
solution is computationally expensive as described in the previous section, we
chose a greedy strategy in the Motley design. In our experimental results
presented later in Section~\ref{section:experiments}, we will show that the
performance of Motley is extremely close to that of the optimal solution.

\subsection{Distance Browsing}
We need to develop an online algorithm that accesses database tuples
(\ie points) incrementally. For this, we adopt the ``distance browsing''
concept proposed in \cite{RTR}, through which it is possible to
efficiently access data points in increasing order of distance from
the query point.  It is predicated on having an index structure with
containment property, such as R-Tree\cite{GUT}, R$^*$-Tree\cite{RSTAR},
LSD-trees\cite{LSD}, etc., built collectively on all dimensions of the
database (more precisely, we need the index to only cover those dimensions
on which point predicates appear in the query workload).  This assumption
appears practical since current database systems such as Oracle, natively
support R-trees~\cite{ORACLE}.  Since the R-tree index is practical only
for low-dimensional data (less than 10 dimensions)~\cite{XTR}, our current
version of Motley is applicable only in such environments. In our future
work, we plan to investigate other indices like the X-tree~\cite{XTR}
which are intended for handling high-dimensional data.

\noindent
\begin{table}[h]
\centerline { \begin{tabular}{|l|}
\hline
{\bf Algorithm NextNearestNeighbor(priority-queue pqueue)}\\
{\bf BEGIN}\\
\hspace*{2mm}{\bf while} (\pqueue is not empty) {\bf do}\\
\hspace*{4mm}   element = get first element of \pqueue\\
\hspace*{4mm}   if (MBRIsPrunable(element)) \hspace*{4mm} continue;\\
\hspace*{4mm}   {\bf if} element is tuple {\bf then} \\
\hspace*{6mm} {\bf return} element\\
\hspace*{4mm}   {\bf else}\\
\hspace*{6mm}     {\bf for} each child c of element {\bf do}\\
\hspace*{8mm}       if (MBRIsPrunable(c)) \hspace*{4mm} continue;\\
\hspace*{8mm}       insert c into \pqueue with key $distance(Q, c)$.\\
\hspace*{6mm}     {\bf done}\\
\hspace*{4mm}   {\bf end-if}\\
\hspace*{2mm}{\bf done}\\
\hspace*{2mm}//complete database scanned, no more tuples left//\\
\hspace*{2mm}{\bf return} null\\
{\bf END}\\
\hline
\end{tabular} }
\caption{Distance browsing algorithm\label{table:NNN}}
\end{table}

To implement distance browsing, a priority queue, \pqueue, is maintained
which is initialized with the root node of the R-Tree. The \pqueue
maintains the objects (R-Tree nodes and data tuples) in increasing order
of distance from query point, that is, the distance of object from the
query point forms the key for that object in the priority queue.

While the distance between a data point and Q is computed in the standard
manner, the distance between a R-tree node and Q is computed as the
minimum distance between Q and any point in the region enclosed by the
MBR (Minimum Bounding Rectangle) of the R-tree node. The distance of
a node from Q is zero if Q is within the MBR of that node, otherwise it
is the distance of the closest point on the MBR periphery. For this,
we first need to compute the distances between the MBR and Q along each
query dimension -- if Q is inside the MBR on a specific dimension, the
distance is zero, whereas if Q is outside the MBR on this dimension, it
is the distance from Q to either the low end or the high end of the MBR,
whichever is nearer. Once the distances along all dimensions are available,
they are combined (based on the distance metric in operation) to get 
the effective distance.

\begin{example} 
Consider an MBR, $M$, specified by ((1,1,1),(3,3,3)) in a 3-D
space.  Let $P_1(2,2,2)$ and $P_2(4,2,0)$ be two data points in this 3-D space.
Then, $spatialdist(M,P_1) = \sqrt{0^2 + 0^2 + 0^2} = 0$ and $spatialdist(M,P_2) =
\sqrt{(4-3)^2 + 0^2 + (0-1)^2} = 1.1414$.

\end{example}

To return the next nearest neighbor, we pick up the first element
of the \pqueue.  If it is a tuple, it is immediately returned as
the nearest neighbor. However, if the element is an R-tree node,
all the children of that node are inserted in the \pqueue.  Note
that during this insertion process, the distance of the object
from the query point is calculated and used as the insertion key.
The insertion process is repeated until we get a tuple as the
first element of the queue, which is then returned.

The above distance browsing process continues until either the diverse result
set is found, or until all points in the database are exhausted, signaled by
the \pqueue becoming empty. The pseudo code for this \emph{NextNearestNeighbor}
algorithm is provided in Table~\ref{table:NNN}. The function MBRIsPrunable is
an optimization and discussed in Section~\ref{section:pruning}.  We assume that
the system has sufficient resources to retain the \pqueue in main memory -- as
shown in our experimental results later, the memory requirements are modest
compared to the capacities of current database servers.

\subsection{Finding Diverse Results}
In the following, we will assume for ease of presentation that we are able
to find fully diverse result sets in the database with regard to the $Q$,
$K$ and \MinDiv specifications.  We refer the reader to \cite{tech-report}
for details about how to handle partially diverse result sets.

\subsubsection{Direct Greedy Approach}
In the Direct Greedy method, we start accessing tuples in increasing order
of distance from query point using the \emph{NextNearestNeighbor} function
discussed above.  The first tuple is always inserted into the result set,
$\cal R$, to satisfy the requirement that the closest tuple to the query
point must figure in the result set.  Subsequently, each new tuple is added
to $\cal R$ if its diversity is greater than \MinDiv with respect to
\emph{all} tuples currently in $\cal R$; otherwise, the tuple is discarded.
This process continues until $\cal R$ grows to contain K tuples.  Note that
the result set obtained by this approach has following property: Let ${\cal
B}={b_1, \ldots, b_K}$ be the sequence formed by any other fully diverse set
such that elements are listed in increasing order of distance from $Q$.  
Now if $i$ is the smallest index such that $b_i \not= R_i \/ ( R_i \epsilon
\cal R) $, then $spatialdist(b_i, Q) \ge spatialdist(R_i, Q)$.  

\begin{figure}[h]
   \centerline{\includegraphics[height=5cm,width=8cm]{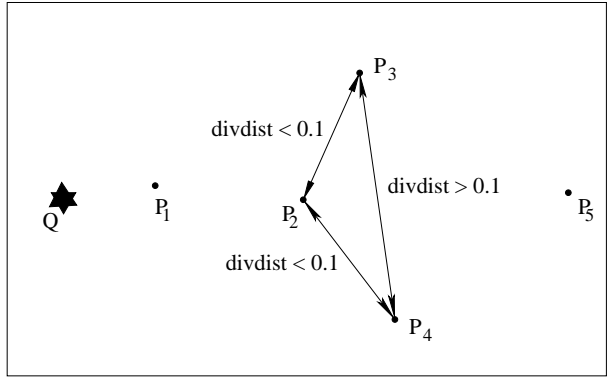}}
   \caption{Poor Choice by Direct Greedy (\MinDiv=0.1)}
   \label{figure:direct}
\end{figure}

While the Direct Greedy approach is straightforward and easy to
implement, there are cases where it may make poor choices as shown
in Figure~\ref{figure:direct}.  Here, Q is the query point, and $P_1$
through $P_5$ are the closest five database tuples.  Let us assume that
the goal is to report 3 diverse tuples with \MinDiv of 0.1. Clearly,
$\{P_1,\;P_3,\;P_4\}$ satisfies the diversity requirement. Also
$DIV(P_1,P_2,V(Q)) = true$. But inclusion of $P_2$ disqualifies the
candidatures of $P_3$ and $P_4$ as both $DIV(P_2,P_3,V(Q)) = false$ and
$DIV(P_2,P_4,V(Q)) = false$. By inspection, we observe that the overall best
choice could be $\{P_1,\; P_3,\; P_4\}$ but Direct Greedy would give
the solution as $\{P_1,\;P_2,\;P_5\}$. Moreover, if point $P_5$ is not
present in the data set, then this approach will fail to return a fully
diverse set even though such a set $\{P_1,\; P_3,\; P_4\}$ is available.

\subsubsection{Buffered Greedy Approach}
To address the above problems, we propose an alternative Buffered Greedy
technique. In this approach, unlike Direct Greedy where at all times we only
retain the diverse points (hereafter called ``leaders'') in the result set, we
maintain with each leader a bounded buffered set of ``dedicated followers'' --
a dedicated follower is a point that is not diverse with respect to a specific
leader but is diverse with respect to \emph{all remaining} leaders. Our
empirical results show that a buffer of capacity $K$ points (where $K$ is the
desired result size) for each leader, is sufficient to produce a near-optimal
solution. The additional memory requirement for the buffers is small for
typical values of $K$ and $D$ (\eg for K=10 and D=10, and using 8 bytes to
store each attribute value, we need only 8K bytes of additional storage).  

Given this additional set of points, we adopt the heuristic that a
current leader point, $L_i$, is \emph{replaced} in the result set by
its dedicated followers $F_i^1, F_i^2, \ldots, F_i^j (j > 1) $ if
(a) these dedicated followers are \emph{all} mutually diverse, and
(b) incorporation of these followers does not result in the premature
disqualification of future leaders.

The first condition is necessary to ensure that the result set contains only
diverse points, while the second is necessary to ensure that we do not produce
solutions that are worse than Direct Greedy. For example, if in
Figure~\ref{figure:direct}, point $P_5$ had happened to be only a little
farther than point $P_4$ such that $DIV(P_2,P_5,V(Q)) = true$, then 
the replacement could be the wrong choice since \{$P_1, P_2, P_5\}$ may turn
out to be the best solution.

In order to implement the second condition, we need to know when it is ``safe''
to go ahead with a replacement \ie we need to know when it is certain that all future
leaders will be diverse from the current set of followers. To achieve this, we take the
following approach: For each point, we consider a hypothetical sphere that
contains all points in the domain space that may have diversity less than
\MinDiv with respect to it. That is, we set the radius $R$ of the sphere to be
equal to the distance of the farthest non-diverse point in domain space.  Note
that this sphere may contain some diverse points as well, but our objective is
to take a conservative approach. Now, the replacement of a leader by a set of
dedicated followers can be done as soon as we have reached a distance greater
than $R$ with respect to the farthest follower from the query point -- this is
because there is no possibility of disqualification beyond this point because
of the appearance of future leaders.

\begin{figure}[h]
   \centerline{\includegraphics[height=6cm,width=9cm]{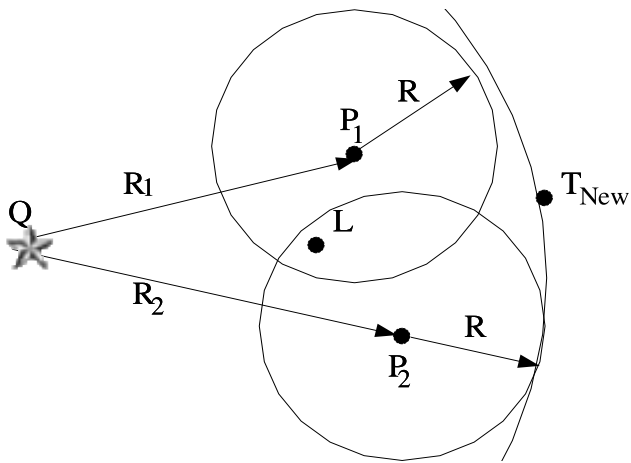}}
   \caption{Heuristic in Buffered Greedy Approach \label{heuristic}}
\end{figure}

For example, in Figure~\ref{heuristic}, the circles around $P_1$ and $P_2$ show
the areas that contain all points that are not diverse with respect to $P_1$
and $P_2$, respectively. Due to distance browsing technique, when we access the
point $T_{new}$ (Figure~\ref{heuristic}), we know that all future points will
be diverse from $P_1$ and $P_2$. At this time, if $P_1$ and $P_2$ are dedicated
followers of $L$ and mutually diverse, then we can replace $L$ by $\{P_1,\;
P_2\}$.

\subsubsection{Integration with Distance Browsing}
The above technique is integrated with the distance browsing approach in the
following manner: For each new tuple returned by the \emph{NextNearestNeighbor}
function, we calculate its distance with respect to Q -- let this distance be
$d_{new}$.  If this tuple has diversity greater than \MinDiv with respect to
all current leaders, then it also becomes a leader. We then immediately
eliminate, from all remaining leaders, their followers who have become
``non-dedicated'' due to the incorporation of the new leader.

Let us consider the alternative situation wherein the new point is not a leader
-- in this case, it is sent to the appropriate leader's buffer if it is a
dedicated follower, otherwise it is discarded.

Now, for each of the original leaders, we select from among the dedicated
followers in their buffer, those points whose distance from the query point is
less than $(d_{new} - R)$. That is, these are the points whose potential
inclusion in the result set will not result in the disqualification of the new
leader.  Among this subset of dedicated followers, we evaluate the largest
group of points that are mutually diverse, and replace the leader with this set
of followers if the group size is greater than one.  When a leader is replaced,
the buffers of all current leaders are visited and those followers which have
now become non-dedicated are removed. The followers in the buffer of the
replaced leader are partitioned into the buffers of the leaders of the new
result set.

While the above computations and reorganizations may appear complex, in
practice they can be completed very quickly because the number of points that
are involved at any given time is fairly small.  The pseudo-code of the
complete Motley algorithm implementing the buffered greedy approach is shown in
Table~\ref{table:MOTLEY}.

\begin{table}[!htbp]
\center{ \begin{tabular}{|l|}
\hline
{\bf Algorithm MOTLEY(Query Q, int K, float \MinDiv)}\\
{\bf BEGIN}\\
\hspace*{2mm} //initialise distance browsing\\
\hspace*{2mm} \pqueue = new priority queue, 
\hspace*{2mm} initialise \pqueue with root node of R$^*$-Tree\\
\hspace*{2mm} let $\cal R = \phi$\\
\\
\hspace*{2mm} {\bf while} (there are less than K leaders in $\cal R$) {\bf do}\\
\hspace*{4mm}   N = NextNearestNeighbor(\pqueue)\\
\hspace*{4mm}   $d_{new}$ = spatialdist(N, Q)\\
\hspace*{4mm}   let L = $\{\; l \; | l \mbox { is leader in } {\cal R}, 
			DIV(l, N, V(Q)) = \mbox { false }\}$\\
\\
\hspace*{4mm}   {\bf if} ($L = \phi$) {\bf then}
\hspace*{10mm}  //N is new leader, remove non-dedicated followers\\
\hspace*{6mm}     {\bf for} each point p in all buffers {\bf do}\\
\hspace*{8mm}       {\bf if} $DIV(p, N, V(Q)) = false$ {\bf then}
\hspace*{4mm}        remove p from buffer
\hspace*{2mm}       {\bf endif}\\
\hspace*{6mm}     {\bf done}\\
\hspace*{6mm}	  $\cal R = \cal R \cup \{ \mbox{N}\}$. Make $N$ as leader in $\cal R$\\
\hspace*{4mm}   {\bf endif}\\
\\
\hspace*{4mm}   //try to find new leaders from dedicated followers\\
\hspace*{4mm}   {\bf for} each leader $l$ of $\cal R$ {\bf do}\\
\hspace*{6mm}     let S = $\{ \; s \; |\; \mbox{s is dedicated followers of } l,
			spatialdist(s, Q) < (d_{new}-R) \}$\\
\hspace*{6mm}     select maximum number of mutually diverse elements from S\\
\hspace*{6mm}     {\bf if} there is more than one element selected {\bf then}\\
\hspace*{8mm}       remove $l$ and make new selected followers as leaders\\
\hspace*{8mm}	    repartition the points in buffers of all leaders in $\cal R$\\
\hspace*{6mm}     {\bf endif}\\
\hspace*{4mm}   {\bf done}\\
\\
\hspace*{4mm}   //add new point into appropriate buffer if it is dedicated follower\\
\hspace*{4mm}   {\bf if} (sizeof(L) $==$ 1) {\bf then}\\
\hspace*{6mm}     let $l$ = element of L\\
\hspace*{6mm}     {\bf if} (buffer of $l$ has free space) {\bf then}
\hspace*{2mm}       add N to buffer of $l$
\hspace*{2mm}     {\bf endif}\\
\hspace*{4mm}   {\bf endif}\\
\hspace*{2mm} {\bf done}  \hspace*{4mm} //while loop\\
{\bf END}\\
\hline
\end{tabular} }
\caption{ Algorithm MOTLEY\label{table:MOTLEY}}
\end{table}

\subsection{Pruning Optimization
\label{section:pruning}}
We now move on to present an optimization through which the processing can be
made more efficient in terms of minimizing the number of database tuples
that are read in arriving at the final result. Note that this optimization
does not affect the contents of the result, but only the \emph{effort} involved
in obtaining this result. The optimization that we propose is the
following: We can prune an MBR (Internal/leaf node of R-Tree) if we are
sure that no point within that MBR can appear in the final result set. 

There are two positions at which pruning can be applied, one when the MBR is
inserted into the \pqueue and the other when it is removed from the \pqueue.
We apply pruning at both times since pruning depends on the contents of the
current result set $\cal R$. Note also that applying pruning before entering
the MBR into the \pqueue reduces the \pqueue size, which can significantly
enhance performance.

\noindent
\begin{table}[ht]
\center { \begin{tabular}{|l|}
\hline
{\bf Algorithm MBRIsPrunable(MBR mbr, result set $\cal R$)}\\
{\bf BEGIN}\\
\hspace*{2mm}let $cnt=0$\\
\hspace*{2mm}{\bf for} each leader l of  $\cal R$ {\bf do}\\
\hspace*{4mm}   farthest = maximum possible diverse point of mbr wrt l\\
\hspace*{4mm}   {\bf if} ($DIV(l, farthest, V(Q)) = false$) {\bf then}\\
\hspace*{6mm}     {\bf if} (l is saturated) {\bf then}
\hspace*{2mm}       {\bf return} true;
\hspace*{6mm}       //no space in buffer to add dedicated followers\\
\hspace*{6mm}     {\bf else} \\
\hspace*{8mm}       $cnt++$
\hspace*{30mm}      //increase the count of non-diverse leaders for this MBR \\
\hspace*{8mm}	    {\bf if} ($cnt>1$) {\bf then} \hspace*{2mm} {\bf return} true;
\hspace*{2mm}     {\bf endif} \hspace*{2mm} //mbr do not contain dedicated followers\\
\hspace*{6mm}     {\bf endif}\\
\hspace*{4mm}   {\bf endif}\\
\hspace*{2mm}{\bf done}\\
\hspace*{2mm}{\bf return} false;
\hspace*{26mm}      //All pruning criteria failed hence mbr can not be prunned \\
{\bf END}\\ 
\hline
\end{tabular}}
{\caption {Algorithm for pruning }\label{table:prune}}
\end{table}
There are two situations under which an MBR can be pruned: Firstly,
we can prune an MBR if all points within it are guaranteed to be
``non-dedicated'' with regard to the current set of leaders. To do this,
we compute for each leader, the point of the MBR that can have maximum
diversity with respect to the leader.  The maximum diverse point is always
a corner of the MBR such that for all dimensions it has maximum possible
distance from the leader.  We calculate the diversity of this maximum
diverse point with respect to each leader and if this diversity is less
than \MinDiv for more than one leader, the complete MBR is pruned.

Secondly, we can also prune as follows: We call a leader to be
\emph{saturated} if its associated buffer is full. For each saturated
leader, we find its maximum diversity with regard to the MBR in the same
manner as described above. If this diversity is less than \MinDiv with
regard to \emph{any} of the saturated leaders, then the MBR can be pruned.

The pseudo code for determining whether an MBR can be pruned is shown
in Table~\ref{table:prune}. It is called in the \emph{NextNearestNeighbor}
function as mentioned previously. 

\subsection{Handling Categorical Attributes
\label{section:categorical}}

In the discussion so far, we had assumed that all attributes are
numeric with inherent ordering among the values. In practice,
however, some of the dimensions may be \emph{categorical} in
nature (\eg \emph{color} in an automobile database), without a natural
ordering scheme. We now discuss how to integrate categorical
attributes into our solution technique. There are two issues
here: 1) how to calculate differences and thereby diversity for
these attributes; and 2) how to incorporate these categorical
attributes in the distance browsing approach.

\subsubsection{Calculating Difference}
In the prior literature, we are aware of two techniques that
address the problem of clustering in categorical spaces -- the
first approach is based on ``similarity''\cite{CCN} and the
second is based on ``summaries''\cite{CAC}. While both techniques
can be used in our framework to calculate diversity, we restrict
our attention to the former in this paper.

The similarity approach works as follows: Greater weight is given to ``uncommon
feature-value matches'' in similarity computations.  For example, consider a
categorical attribute whose domain has two possible values, $a$ and $b$.  Let
$a$ occur more frequently than $b$ in the dataset. Further, let $i$ and $j$ be
tuples in the database that contain $a$, and let $p$ and $q$ be tuples that
contain $b$.  Then the pair $p,q$ is considered to be more similar than the
pair $i,j$, \ie $Similar(i, j) > Similar(p, q)$;  in essence, tuples that match
on less frequent values are considered more similar.

Quantitatively, similarity values are normalized to the range [0,1]. The
similarity is zero if two tuples have different values for the categorical
attribute.  If they have the same value $v$, then the similarity is computed as
follows:
\[ 
Sim(v) = 1 - \sum_{l \in MoreSim(v)} \frac {f_l(f_l-1)}{n(n-1)}\\ 
\] 
where $f_l$ is frequency of occurrence of value $l$, $n$ is the number of
tuples in the database, and \emph{MoreSim(v)} is the set of all values in
the categorical attribute domain that are more similar or equally similar
as the value $v$ (\ie they have lesser frequency).

We cannot directly use the above formula in our diversity framework since
our goal is to measure \emph{difference}, not similarity. At first glance,
the obvious choice might seem to be to set \emph{difference}$(\delta)=1-$
\emph{similarity}.  But this has two problems: Firstly, tuples with
different values in the categorical attribute will have a difference of 1.
Secondly, tuples with identical values will have a non-zero difference.
Both these contradict our basic intuition of diversity.

Therefore, we set the definition of difference as follows: If two tuples
have the same attribute value, then their difference is zero. Tuples with
different values will have difference based on the frequencies of their
attribute values. The more frequent the values, the more is the difference.
For example, if the categorical attribute has values $a$, $b$ and $c$ in
decreasing order of frequencies, $\delta(a,c) > \delta(b,c)$, since $a$ is more
frequent than $b$. In general, given points with categorical attribute
values $v_1$ and $v_2$, we can quantitatively define 
\begin{eqnarray*}
\delta({v_1,v_2}) &=& 1 - Sim(v_1) * Sim(v_2) \hspace*{5mm}\mbox{ if } v_1 \neq v_2 \\
		&=& 0  \hspace*{41mm}	\mbox{ if } v_1 = v_2
\end{eqnarray*}

\subsubsection{Integrating with Distance Browsing}
To integrate categorical attributes with distance browsing, a
prerequisite is a containment index structure that can handle
categorical attributes. This can be achieved using
recently-proposed indexes such as the M-tree~\cite{MTR} or the
ND-Tree~\cite{NDT}, which specifically provide this functionality
on categorical attributes.

\subsection{Partially Specified Queries}
\label{secpartialquery}
Upto this stage, we had assumed that an R-Tree on exactly the set of point
attributes mentioned in the query is available. But, in general, this need
not be the case since the user query may involve only a subset of
attributes on which the R-tree was built. Such queries are called
\emph{partially specified queries}~\cite{TOPK}. One option is to to build
R-Trees on all possible attribute combinations but this is infeasible in
practice due to the large number of combinations.  Therefore, we instead
initially build a R-Tree on all the possible query dimensions and, for
partially-specified queries, take a (logical) projection of the
R-tree on the associated sub-space.A problem with this approach is that
when the number of attributes in the partially-specified query is only a
few, the performance of the R-tree projection may deteriorate due to the
increased overlap in MBRs. We quantitatively assess this issue in our
experimental study.

\subsection{Nearest Neighbor queries}
As mentioned earlier, Motley can also be used to execute nearest
neighbor queries simply by setting \MinDiv$=0$. Further, it does this as
efficiently as the direct distance browsing technique for KNN, without
adding any additional overheads.   A detailed evaluation of distance
browsing as compared to more traditional approaches for KNN has been
done in ~\cite{RTR} -- their results indicate that distance browsing
outperforms traditional KNN.  Therefore, Motley can be seamlessly used
for both the KNN and KNDN problems.

Another point to note is that some applications may want their diversity
limited only to the elimination of \emph{exact duplicates}.  This can be easily
achieved by setting \MinDiv to a very small but non-zero value.

\section{Experiments
\label{section:experiments}}

We conducted a variety of experiments to evaluate the quality and
efficiency of the Motley algorithm with regard to producing a
diverse set of answers. In this section, we describe the
experimental framework and the results.

We used three datasets in our experiments, representing a combination
of real and synthetic data, similar to those used in ~\cite{TOPK}.
Dataset 1 is a projection of the US Census Bureau data~\cite{DATA1},
containing 32,561 tuples. The original dataset contained 14 attributes,
from which we projected four numerical valued attributes, representing
\emph{age}, \emph{wage}, \emph{education}, and \emph{hours of work per
week}.  Dataset 2 is a projection of another real data set~\cite{DATA2}
containing 581,012 tuples and 54 attributes related to geographic terrain
data.  In this dataset also we selected 4 numerical valued attributes,
representing Elevation, Aspect, Slope, and Distance.  Finally, Dataset
3 is synthetically generated data that follows Zipf~\cite{zipf}
distribution. For generating the data, a Zipf parameter of 1.0 is used for
all attributes, resulting in highly skewed data. This dataset contains
50,000 tuples over six dimensions.

The majority of our experiments involve uniformly distributed point
queries across the whole data space, with the attribute domains
in all datasets normalised to the range [0,1].  We consider both
fully-specified point queries, that is, queries over all dimensions of
the data, as well as partially-specified point queries, wherein only a
subset of the dimensions appear in the query. The default value of K,
the desired number of answers, was 10, unless mentioned otherwise, and
\MinDiv was varied across the [0,1] range.  In practice, we would expect
that \MinDiv settings would be on the low side, typically not more than
0.2, and we therefore focus on this range in most of our experiments.
The decay rate ($a$) of the weights in Equation~\ref{eq:weights} was
set to 0.1 for all experiments.

Our results were obtained on a Pentium-III 800 MHz machine with
128MB of main memory and running Linux 7.2. The R-tree
(specifically, the R$^*$ variant~\cite{RSTAR}) was created with a
fill factor of 0.7 and branching factor 64, using the source code
available from \cite{CODE}. Ten buffers, each of size 4 KB, were
assigned to the R$^*$ tree, with random replacement policy (since
no node is scanned twice in Motley, the replacement policy is not
an issue). The disk occupancy of the R$^*$ tree for the three
datasets was 4.1MB, 79MB, and 8.6MB, respectively. For buffered
greedy approach, the number of buffers for each leader was set
equal to K, representing a maximum memory storage requirement
which was of the order of a few kilobytes for all
the datasets.

We measured the quality of our solution using the scoring metric
of Equation~\ref{eqn:score}, with a Euclidean distance function
for measuring distances, and the harmonic mean for computing
the aggregate distance.  We also measured the average distances
of points in the result set.  The efficiency of our algorithm was evaluated by
counting the number of tuples read from the database. This
measure includes the tuples that need to be read for insertion
into the priority queue.  Since the in-memory processing is
relatively very quick, the disk activity indicated by the number
of tuples read forms a reasonable metric.

We also report the percentage of time required in our approach
with respect to the \emph{sequential scan} method, which essentially
represents an \emph{upper bound} on performance.  In the sequential scan
method, we read all tuples(points), sort them based on their
distance from the query point, and then make one pass to determine the
diverse result set using the Buffered Greedy technique.

Finally, the worst-case main-memory usage across all queries for
each dataset was measured, and we found that all dynamic
structures including the priority queue could be accommodated
within 10 MB for Datasets 1 and 3, whereas it was around 40 MB
for Dataset 2.

In the remainder of this section, we initially present results
for fully-specified queries and subsequently for partially-specified
queries. 
\subsection{Result-set Characteristics}
\label{secquality}

\begin{figure*}[t]
   \includegraphics[height=2.2in,width=6.0in]{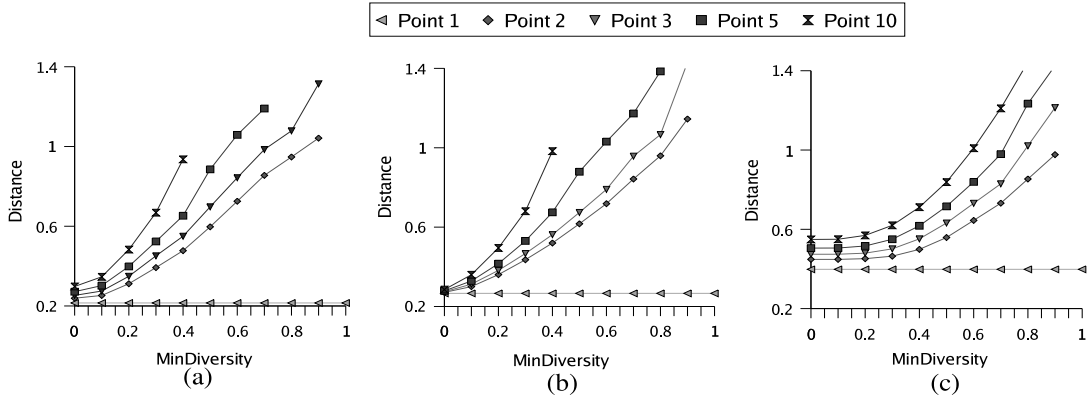}
   \caption{Average distance of diverse points (a) Dataset 1 (b) Dataset 2 (c) Dataset 3}
   \label{avgdist}
\end{figure*} 

In Figure~\ref{avgdist}, the average distances of the \emph{diverse} points as
a function of \MinDiv are shown for all three datasets. For the sake of graph
clarity, the distances are shown only for the 1st, 2nd, 3rd, 5th, and 10th (\ie
last) diverse points -- the behavior of the other points was similar.  Note
that, the result set $\cal R$ may not be fully-diverse at higher values of
\MinDiv. The non-diverse points are excluded while computing the averages in
Figure~\ref{avgdist}.  The reason that the curves terminate early without going
across the entire \MinDiv range is because at higher values of \MinDiv, there
may be no queries possible for which a particular $k^{th}$ point can be
diverse. These graphs also show the maximum \MinDiv setting for which a fully
diverse result set of size 10 can be found.

We see in Figure~\ref{avgdist} that the distance of the first point is
independent of \MinDiv  -- this is an artifact of our requirement that the
closest point to the query should always form part of $\cal R$ and is
therefore not impacted by the \MinDiv setting.  Secondly, comparing
Figures~\ref{avgdist}(a) and ~\ref{avgdist}(c), whose datasets are 4 and
6-dimensional, respectively, we observe that the result set on
6-dimensional data becomes partially diverse at higher values of \MinDiv
than the 4-dimensional data. This is expected because as the number of
dimensions increases, the distance between individual points tends to
increase, and therefore the diversity also increases.

\begin{figure}[ht]
   \centerline{\includegraphics[height=2.2in,width=6.5in]{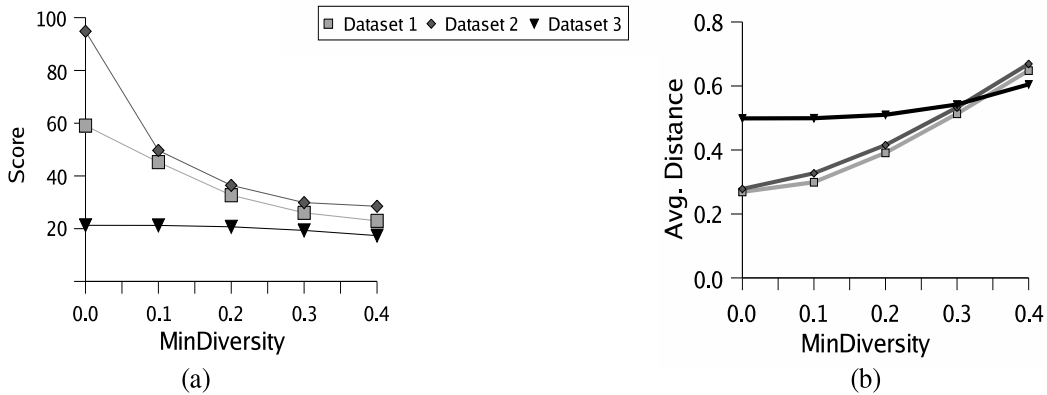}}
   \caption{(a)Score of diverse set (b)Average distance of all points
	\label{characteristics}}
\end{figure}

Figure~\ref{characteristics}(a) shows the scores of the result set $\cal
R$ for the three different datasets. With increase in \MinDiv,
the score of the result set decreases as expected because of the increase in
the distance of points and hence their harmonic mean value.
Figure~\ref{characteristics}(b) shows the average distance of the points
in $\cal R$ for the three datasets.  It shows the cost to be paid in
terms of distance in order to obtain result diversity. The important point to
note here is that in all the three datasets, for values of \MinDiv up to
0.2, the distance increase is marginal. Since we expect that users will
typically use \MinDiv values between 0 and 0.2, it means that diversity
can be obtained at relatively little cost in terms of distance. 

Note that that the graphs for Dataset 3 are almost flat in both
Figures~\ref{characteristics}(a) and ~\ref{characteristics}(b).  This is
because of the high skew along each dimension in this dataset and the
uniformly distributed query workload, which makes the spatially-nearest
neighbors themselves to be diverse for most of the queries.

\begin{figure}[ht]
   \centerline{\includegraphics[height=5cm,width=8cm]{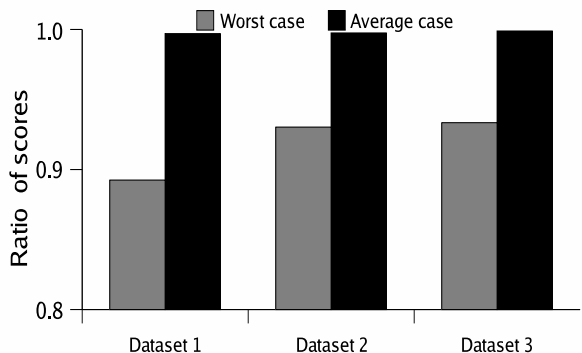}}
   \caption{MOTLEY vs. Optimal}
   \label{ratiohm}
\end{figure}

We now move on to characterizing the quality of the result set
provided by Motley, which is a greedy online algorithm, against
an optimal off line algorithm.  This performance perspective is
shown in Figure~\ref{ratiohm} which presents the average and
worst case ratio of the result set scores for all three datasets.
As can be seen in figure, the average case is almost optimal
(note that the Y-axis of the graph begins from 0.8), indicating
that Motley typically produces a \emph{close-to-optimal}
solution. Further, even in the worst-case, the difference is only
around ten percent. We further investigated this issue by evaluating, for
the cases where the Motley and optimal were different, the
percentage of points that were common between the Motley result
set and the optimal-set. Our experimental results showed that
more than 90 percent of the answers were common. That is, even in
the cases where a sub-optimal choice was made, the errors were
mostly restricted to \emph{one or two} points, out of the total
of ten answers.

\subsection{Execution Efficiency}
\label{secperformance}
\begin{figure*}[t]
   \centerline{\includegraphics[height=6cm,width=16cm]{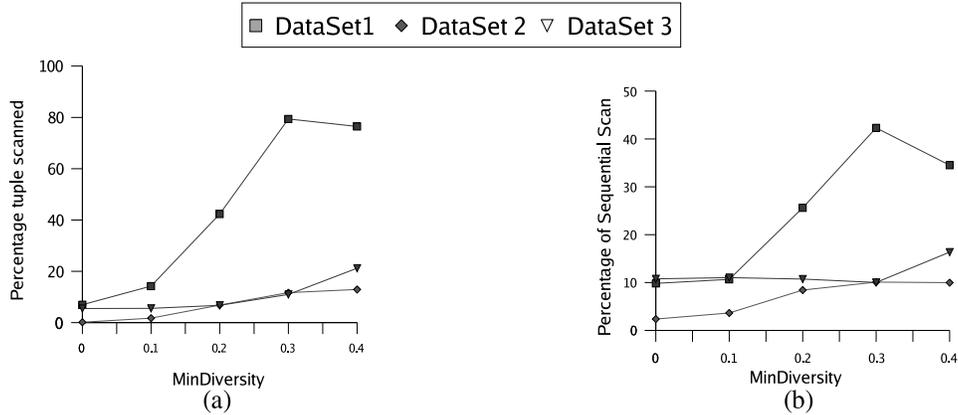}}
   \caption{(a) Average number of tuples read \hspace*{0.5in}
            (b) Percentage of sequential scan \label{perfom2}}
\end{figure*}

Having established the high-quality of Motley answers, we now
move on to evaluating its execution efficiency. In
Figure~\ref{perfom2}(a), we show the average fraction of tuples
read to produce the result set as a function of \MinDiv for
all three datasets.  We see here that at lower values of
\MinDiv, the number of tuples read are small because we obtain
$K$ diverse tuples after processing only a small number of points.
At higher values of \MinDiv, the number of MBRs pruned are
more and hence tuples read are less. For example, with Dataset 2, we always read
less than 14\% of the total tuples.

An important point to note here is that the performance of the traditional
KNN search, obtained by setting \MinDiv = 0, is extremely good, indicating
that Motley is a general algorithm that does not sacrifice performance on
traditional KNN search in order to accommodate diversity goals. To further
confirm this, we compared the performance of MOTLEY with respect to the
reported performance of the Dyn KNN algorithm\cite{TOPK} 
for Datasets 1 and 2, with $K=100$. While
for the small-sized Dataset 1, the number of tuples read by Dyn and MOTLEY
were nearly the same, for the comparatively larger Dataset 2, the number
of tuples read by Dyn was nearly 2\% whereas MOTLEY reads only 0.5\%.

Figure~\ref{perfom2}(b) shows the percentage of time required by Motley with
respect to sequential scan for different values of \MinDiv for all three
datasets. From the graph, it can be seen that as data size increases
Motley requires a progressively smaller percentage of time than sequential
scan.  Even with Dataset 1, which is a rather modest 32,561 tuples in size,
Motley requires less than 40\% of the time taken by sequential scan.
When we consider the much larger Dataset 2, which contains more than
half a million tuples, Motley consistently takes only about 10\% of the
sequential scan time.  Further, note that the numbers reported here
are \emph{conservative} since the sorting of the dataset required by
sequential scan was done in memory by allocating sufficient resources --
in the general case, however, external sorting would have to be carried
out, and the performance of sequential scan would become even worse.

\begin{figure}[h]
   \centerline{\includegraphics[height=5cm,width=7.5cm]{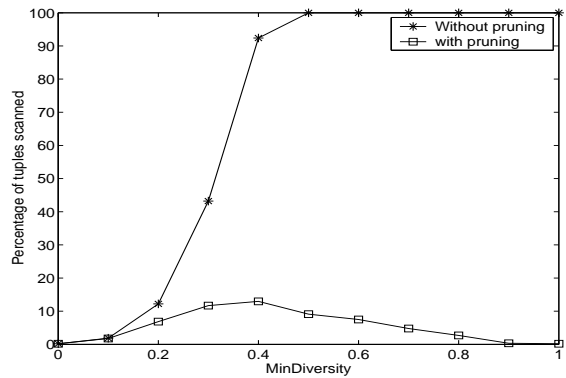}}
   \caption{Impact of Pruning}
   \label{noprune}
\end{figure}

To quantify the impact of pruning, we ran Motley with and without
the pruning optimizations.  Figure~\ref{noprune} shows a sample
performance on Dataset 1 with K=10 and default settings.  We see
here that a substantial improvement is produced by the inclusion of
these pruning optimizations.

\subsection{Effect of K}
\label{seceffectofk}

\begin{figure*}[t]
   \includegraphics[height=5cm,width=16cm]{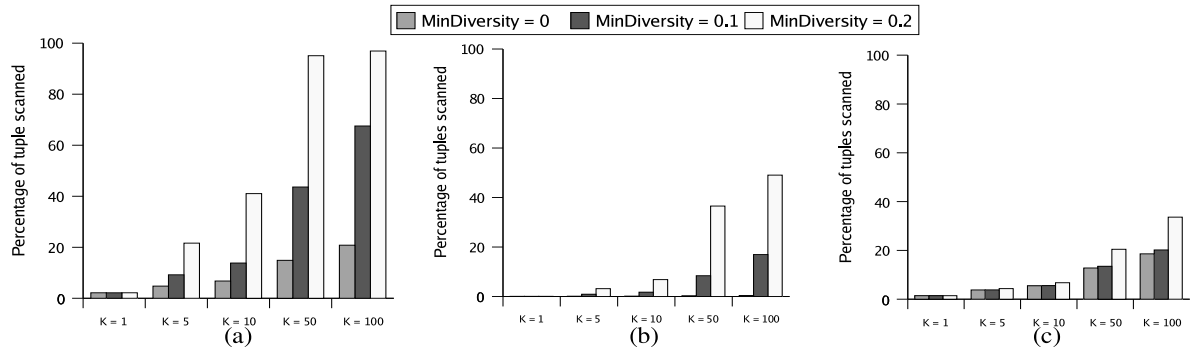}
   \caption{Average number of tuples read for different 
	values of K  (a) Dataset 1 (b) Dataset 2 (c) Dataset 3}
   \label{kon}
\end{figure*}

This experiment evaluates the effect of $K$, the number of
answers, on the algorithmic performance.
Figure~\ref{kon} shows the percentage of tuples read as a function of \MinDiv
for different values of K ranging from 1 to 100. For $K=1$, it is equivalent to
the traditional NN search, irrespective of \MinDiv, due to requiring the
closest point to form part of the result set. As the value of $K$ increases, the
number of tuples read also increases, especially for higher values of \MinDiv.
However, we can expect that users will specify lower values of \MinDiv for
large $K$ settings.
Dataset 1 contains only 32561 tuples so the size of the MBR represented
by each R-Tree node is too large for pruning to be effective. Therefore, at 
$K=50$ and $K=100$, almost the entire database is scanned.

\subsection{Partially-specified Point Query}
\begin{figure*}[t]
   \centerline{\includegraphics[height=6cm,width=16cm]{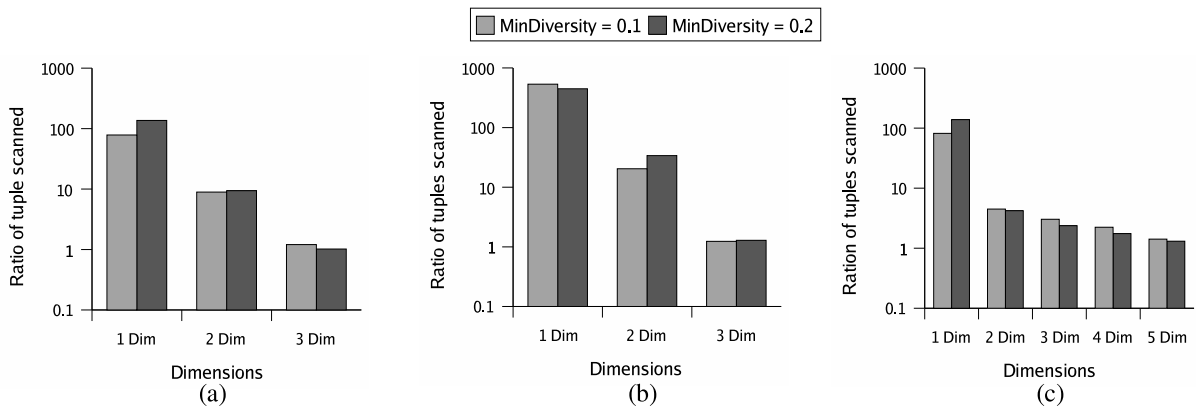}}
   \caption{Ratio of tuples read for partially-specified queries
	(a)Dataset 1 (b)Dataset 2 (c)Dataset 3 }
   \label{subdim}
\end{figure*}
We now move on to evaluating the performance when the point attributes in
the query specify
only a \emph{subset} of the database dimensions. An issue here is whether we
should build an R-tree specifically on the point attributes or should
we simply project the global R-tree built across all dimensions onto
the point attributes.  Ideally, we would like to have only a single
global R-tree, since building R-trees for all possible combinations of
attributes ($2^D$) is prohibitively expensive, as mentioned earlier in
Section~\ref{section:motley}. We measure the performance impact of these
alternative choices here.

Figure~\ref{subdim} shows the ratio of tuples read for a global
R-tree as opposed to a customized R-tree for all three data sets, as a
function of the number of point attribute, for \MinDiv
settings of 0.1 and 0.2.  We see here that while reduction of one
or two dimensions does not unduly increase the number of tuples
read, the performance degrades substantially at lower dimensions
-- this is because the \emph{overlap} among MBRs in the global R-tree
becomes high at these low dimensions.

This suggests that if the query workload has a large number of
low-dimensional queries, it may be worthwhile to build a 
carefully chosen hierarchy of R-trees, some catering to the low
dimensions and others catering to the higher dimensions.

\label{secpartquery}

\section{Related Work
\label{section:related}}

To the best of our knowledge, the KNDN problem investigated in this paper
has not appeared elsewhere in the literature.  It has its roots in the KNN
problem, sometimes referred to as Top-K, that has been extensively studied
in the last decade -- we refer the reader to \cite{TOPK,RTR} for recent
surveys of this literature.  The two major trends in this corpus is one
based on computing nearest K using standard database statistics
(\eg\cite{TOPK}), while the other is based on spatial indices such as the
R-tree (\eg\cite{RTR,KNN}) -- we have used the latter approach in this
paper.

It may appear that an alternative solution to the KNDN problem would be to
initially process the data into clusters using algorithms such as BIRCH
\cite{BIRCH}, replace all clusters by their representatives, and then to
apply the traditional KNN approach on this summary database. There are two
problems here: Firstly, since the clusters are pre-determined, there is
no way to dynamically specify the desired diversity, which may vary from
one user to another or may be based on the specific application that is
invoking the KNDN search.  Secondly, since the query attributes are not
known in advance, we potentially need to do clustering in each subspace
of dimensions, which may become infeasible due to the exponential number
of such subspaces.  Finally, this approach cannot provide the traditional
KNN results.

Yet another approach to produce a diverse result set could be to run
the standard KNN algorithm, cluster its results, replace the clusters
by their representatives, and then output these representatives as the
diverse set.  The problem with this approach is that it is not clear what
the original number of required answers  should be set to such that there
are finally K representatives.  If the original number is set too low,
then the search process has to be restarted, whereas if the original
number is set too high, a lot of wasted work ensues.

Finally, the Skyline operator~\cite{SKL} implements a different notion of
``interesting'' tuples in that, given a database of tuples, it finds the
\emph{dominating} tuples among this set. A tuple $T_1$ is said to dominate
$T_2$ if $T_1$ is superior to $T_2$ in all attributes with respect to given
query point. The skyline operator returns the set of tuples that are not
dominated by any of the other tuples in the database. It is different from
our approach in that the notion of domination is not with respect to a
query point, and further it is possible that the results may be highly
clustered spatially, resulting in very low diversity from our perspective.

\section{Conclusions}
\label{section:conclusions}

In this paper, we introduced the problem of finding the K Nearest Diverse
Neighbors (KNDN), where the goal is to find the closest set of answers such that the
user will find each answer sufficiently different from the rest, thereby adding
value to the result set.  We provided a quantitative notion of diversity that
ensured that two tuples were diverse if they differed in at least one dimension
by a sufficient distance, and presented a two-level scoring function to
integrate the orthogonal notions of distance and diversity.

We presented MOTLEY, an online algorithm for addressing the KNDN problem,
based on a greedy approach integrated with a distance browsing technique. A
buffered variation of Motley was introduced to improve the solution quality
of the basic greedy approach, while pruning optimizations were incorporated
to improve the runtime efficiency.  Our experimental results with a variety
of real and synthetic data-sets demonstrated that Motley can provide
high-quality diverse solutions at a low cost in terms of both result
distance and processing time.  In fact, Motley's performance was close
to the optimal in the average case and only off by around ten percent in
the worst case.

In our future work, we plan to extend our implementation of Motley to
handle categorical attributes in accordance with the mechanisms discussed
in this paper.
 
\bibliographystyle{latex8}

\small
\baselineskip 10pt

\begin{thebibliography}{99}

%
%
%

\bibitem{RSTAR} N. Beckmann, H. Kriegel, R. Schneider and B. Seeger,
{\it The R$^*$-tree: An efficient and robust access method for points and rectangles},
Proc. of ACM SIGMOD Intl. Conf. on Management of Data, 1990.

\bibitem{XTR} S. Berchtold, D. Keim and H. Kriegel,
{\it The X-tree: An Index Structure for High-Dimensional Data},
Proc. of $22^{nd}$ Intl. Conf. on Very Large Data Bases, 1996. 

\bibitem{SKL} S. Borzsonyi, D. Kossmann and K. Stocker,
{\it The Skyline Operator},
Proc. of $17^{th}$ Intl. Conf. on Data Engineering, 2001.

\bibitem{GINI} L. Breiman, J. Friedman, R. Olshen and C. Stone,
{\it Classification and Regression Trees },
Chapman and Hall, 1984.

\bibitem{GOWER} J. Gower,
{\it A general coefficient of similarity and some of its properties},
Biometrics 27, 1971.

%
%
%

\bibitem{TOPK} N. Bruno, S. Chaudhuri and L. Gravano,
{\it Top-K Selection Queries over Relational Databases: 
Mapping Strategies and Performance Evaluation},
ACM Trans. on Database Systems, 27(2), 2002.

\bibitem{MTR}  P. Ciaccia, M. Patella and P. Zezula,
{\it M-Tree: An Efficient Access Method for Similarity Search 
in Metric Spaces},
Proc. of $23^{rd}$ Intl. Conf. on Very Large Data Bases, 1997.

\bibitem{CAC} V. Ganti, J. Gehrke and R. Ramakrishnan,
{\it CACTUS-Clustering Categorical Data using Summaries},
Proc. of ACM Knowledge and Data Discovery Conf., 1999.

\bibitem{MG} M. Grohe, 
{\it Descriptive and Parameterized Complexity},
Computer Science Logic, 13th Workshop, number 1683 in Lecture
Notes in Computer Science, pages 14-31. Springer-Verlag,
September 1999.

\bibitem{GUT} A. Guttman,
{\it R-trees: A dynamic index structure for spatial searching},
Proc. of ACM SIGMOD Intl. Conf. on Management of Data, 1984.

\bibitem{LSD} A. Henrich,
{\it The LSD-tree: An Access Structure for Feature Vectors},
Proc. of $14^{th}$ Intl. Conf. on Data Engineering, 1998. 

\bibitem{RTR} G. Hjaltason and H. Samet,
{\it Distance Browsing in Spatial Databases},
ACM Trans. on Database Systems, 24(2), 1999. 

\bibitem{tech-report} A. Jain, P. Sarda and J. Haritsa, 
{\it Providing Diversity in K-Nearest Neighbor Query Results},
Tech. Report, DSL, Indian Institute of Science, 2003.

%
%
%

\bibitem{ORACLE} R. Kothuri, S. Ravada and D. Abugov, 
{\it Quadtree and R-tree indexes in Oracle Spatial: A comparison using GIS data},
Proc. of ACM SIGMOD Intl. Conf. on Management of Data, 2002.

\bibitem{CCN} C. Li and G. Biswas,
{\it Conceptual Clustering with Numeric and Nominal Mixed Data -- A New Similarity Based System},
IEEE Trans. on Knowledge and Data Engineering, 14(4), 2002. 

\bibitem{mont}\emph{www.elibronquotations.com}

\bibitem{NDT} G. Qian, Q. Zhu, Q. Xue and S. Pramanik,
{\it The ND-Tree: A Dynamic Indexing Technique for Multidimensional
Non-ordered Discrete Data Spaces},
Proc. of $29^{th}$ Intl. Conf. on Very Large Data Bases, 2003.

\bibitem{KNN} N. Roussopoulos, S. Kelley and F.Vincent,
{\it Nearest Neighbor Queries},
Proc. of ACM SIGMOD Intl. Conf. on Management of Data, 1995.

%
%
%

%
%
%

\bibitem{BIRCH} T. Zhang, R. Ramakrishnan and M. Livny,
{\it BIRCH: An Efficient Data Clustering Method for Very Large Databases},
Proc. of ACM SIGMOD Intl. Conf. on Management of Data, 1996.

\bibitem{zipf} G. Zipf. {\it Human behaviour and the principle of least effort.} Addison-Wesley, 1949.

%
%
%

\bibitem{DATA1}	ftp://ftp.ics.uci.edu/pub/machine-learning-databases/census-income
\bibitem{DATA2}	ftp://ftp.ics.uci.edu/pub/machine-learning-databases/covtype
\bibitem{CODE}  http://www.cs.ucr.edu/marioh/spatialindex/	
\end{thebibliography}

\end{document}